\documentclass[pof,aps,preprint,onecolumn]{revtex4}
\usepackage{graphicx}
\def\strutdepth{\dp\strutbox}
\def\nw#1{\strut\vadjust{\kern-\strutdepth\vtop to0pt{\vss\hbox to\hsize
{\hskip\hsize\hskip5pt$\leftarrow$\hss\strut}}}{\em #1}}
\begin{document}

\title{Maximum speed of dewetting on a fiber}

\author{ Tak Shing Chan, Thomas Gueudr\'e, and Jacco H. Snoeijer}
\affiliation{
Physics of Fluids Group and J. M. Burgers Centre for Fluid Dynamics,
University of Twente, P.O. Box 217, 7500 AE Enschede, The Netherlands \\
}

\date{\today}

\begin{abstract}
A solid object can be coated by a nonwetting liquid since a receding contact line cannot exceed a critical speed. We theoretically investigate this forced wetting transition for axisymmetric menisci on fibers of varying radii. First, we use a matched asymptotic expansion and derive the maximum speed of dewetting. For all radii we find the maximum speed occurs at vanishing apparent contact angle. To further investigate the transition we numerically determine the bifurcation diagram for steady menisci. It is found that the meniscus profiles on thick fibers are smooth, even when there is a film deposited between the bath and the contact line, while profiles on thin fibers exhibit strong oscillations. We discuss how this could lead to different experimental scenarios of film deposition.
\end{abstract}

\maketitle

\section{Introduction}

A convenient way to deposit a thin liquid layer on a surface is by withdrawing a solid from a liquid reservoir. The film is dragged along with the solid due to the viscous friction of  the liquid. This principle is known as dip-coating and is a commonly used technique in industrial contexts~\cite{Q99,WeRu04}. Once deposited on the surface, the film often has a thickness as predicted by Landau, Levich~\cite{LL42} and Derjaguin~\cite{D43}, scaling with speed $U$ of withdrawal as $h \propto U^{2/3}$. Recently, however, a different class of solutions were identified, which are much thicker and scale as $h~\propto U^{1/2}$~\cite{SZAFE08}. These thick films were indeed realized experimentally in the case where the solid was partially wetting.

\begin{figure}[t!]
\includegraphics[width=8.0cm]{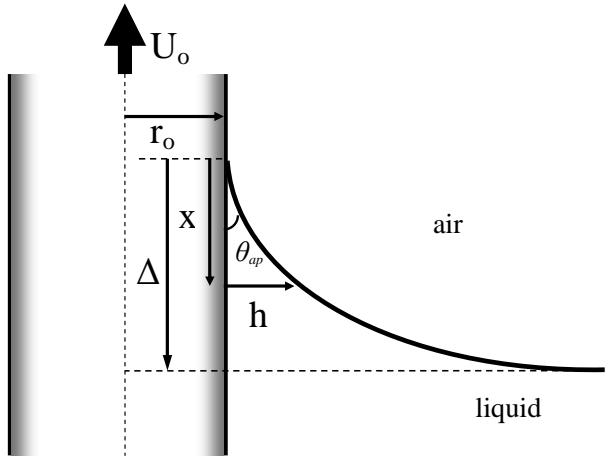}
\caption{Schematic representation of the dip-coating setup: A fiber or cylinder of radius $r_0$ is withdrawn with speed $U_0$ from a bath of viscous liquid. The axisymmetric meniscus profile is characterized by $h(x)$, while $\Delta$ is denotes the maximum rise above the reservoir.}
\label{fig.sketch}
\end{figure}

The conditions of partial wetting introduces another interesting feature, namely that the film entrainment only appears above a critical velocity of withdrawal~\cite{V76,C86,G86,SRC05,BEIMR09}. Below this speed the contact line finds at a steady position, indicated as the meniscus rise $\Delta$ (Fig.~\ref{fig.sketch}). Due to viscous drag between the liquid and the solid, the dynamical position of $\Delta$ is higher than at equilibrium. This means that the apparent contact angle $\theta_{ap}$ of the dynamical meniscus is smaller than the equilibrium angle $\theta_e$. The simplest interpretation of the transition to film deposition is that the apparent contact angle $\theta_{ap} \rightarrow 0$ at some critical plate velocity.
This idea was already postulated by Derjaguin and Levi~\cite{DL64}, although the energy argument given by de Gennes~\cite{G86} suggested a nonzero $\theta_{ap}$ at the transition. The hypothesis of $\theta_{ap}$=0, however, was given a rigorous mathematical basis (for a flat solid) by asymptotic expansions of the lubrication equations~\cite{E04b,E05}. Actually, it was shown by~\cite{BEIMR09} that de Gennes energy argument can be extended to incorporate interface curvature: this exactly gives the lubrication equation, meaning that also the energy argument leads to a zero $\theta_{ap}$ at the transition. This theory gives a simple prediction for the maximum rise, based on the static meniscus solution with vanishing contact angle -- for a fiber of radius $r_0$ this simply becomes~\cite{LL84a,DFJ74}

\begin{eqnarray}\label{eq:dmax}
\Delta_{\rm max} \simeq
\left\{
\begin{array}{l l}
r_0 (\ln \frac{4 \ell_c}{r_0}-c) & \quad \textrm{for $r_0 \ll \ell_c$} \\
\sqrt{2} \ell_c &  \quad \textrm{for $r_0 \gg \ell_c$}.
\end{array}
\right.
\end{eqnarray}
Here $\ell_c=(\gamma/\rho g)^{1/2}$ is the capillary length based on surface tension $\gamma$, density $\rho$, gravity $g$ and $c$ is Euler's constant (0.57721). At intermediate radii $r_0 \sim \ell_c$, the maximum rise can be determined numerically.

Experimentally, the description of the forced wetting transition has remained ambiguous. The condition of a vanishing apparent contact angle was convincingly shown by Sedev \& Petrov~\cite{SP91}. When withdrawing fibers or thin cylinders ($r_0/\ell_c \sim 0.06-1$), they found a maximum rise of the meniscus consistent with (\ref{eq:dmax}). Using cylinders of larger radii ($r_0/\ell_c \sim 10$), Maleki \emph{et al.}~\cite{MRQG07} found zero or nonzero $\theta_{ap}$ at the transition, depending on the way $\theta_{ap}$ was determined. When using the criterion based on the meniscus height, the transition was found slightly before reaching $\Delta_{\rm max}$. Yet another set of experiments using a flat plate ($r_0/\ell_c=\infty$) displayed a transition to film deposition clearly before reaching the maximum rise~\cite{SDAF06,DFSA07}. Still, during the unsteady entrainment phase the maximum recorded speed was reached exactly at $\sqrt{2}\ell_c$. Note that in these experiments, the deposited liquid was not simply the Landau-Levich-Derjaguin film, but gave rise to thick films and even shock solutions. It was argued that the presence of these dynamical solutions are related to the pre-critical onset of entrainment~\cite{SADF07}, but an explanation is still lacking.

An additional complexity is that the contact line can spontaneously develop sharp corner structures, or even zig-zags. This has been observed in dip-coating~\cite{BR79}, splashing~\cite{DYCB07}, immersion lithography~\cite{WPER11,BKST1} and for drops sliding down an inclined plane~\cite{PFL01,TNBPV02}. The conical structure of the interface near the contact line renders the problem truly three-dimensional, which affects the balance of the capillary forces~\cite{LS03}. For sliding drops, it has been observed experimentally and described by a 3D lubrication model, that this change in geometry indeed leads to a nonzero apparent contact angle at the transition to liquid deposition~\cite{LDL05,SLLSE07}. This raises the question of how the geometry of the flow  can influence the critical speed of wetting~\cite{ZSE09}.

In this paper we theoretically study the withdrawal of fibers of arbitrary radii. By varying the ratio $r_0/\ell_c$, we continuously cover the range from thin fibers to the flat plate. First, we extend the asymptotic analysis that was previously done for the flat plate~\cite{E04b,E05} to the limit of thin fibers (Sec.~\ref{sec:asymptotics}). To resolve the singularity of viscous stress near the contact line~\cite{HS71,H83}, we introduce a slip length $\lambda$~\cite{LBS05,CCSC05}. Other types of microscopic regularization will give similar results~\cite{BEIMR09}. Typical values for the slip and capillary lengths are $\lambda \sim 10^{-9}$m and $\ell_c \sim 10^{-3}$m respectively. We can thus exploit the hierarchy of length scales

\begin{equation}
\lambda \ll r_0 \ll \ell_c,
\end{equation}
and perform a matched asymptotic expansion. The control parameter is the capillary number ${\rm Ca}=U_0\eta/\gamma$, which is the speed of withdrawal scaled by viscosity $\eta$ and surface tension $\gamma$. The analysis yields the critical capillary number, which depends on the value of $r_0$, and confirms that the maximum speed coincides with $\theta_{ap}=0$, for all fiber radii $r_0$. In this sense, the change in geometry does not qualitatively change the nature of the critical point. However, striking differences do show up when computing numerically the complete bifurcation diagrams for all steady solutions (Sec.~\ref{sec:bifurcation}). These diagrams include solution branches above $\Delta_{\rm max}$ that are unstable, but which have been observed as transients during film deposition for the plate case~\cite{DFSA07}. We find that for small fiber radii much below $\ell_c$, the steady solutions no longer smoothly join the film solutions that mediate the deposition. In the Discussion section we speculate that this is why, experimentally, it is easier to approach the critical point for thin fibers (Sec.~\ref{sec:discussion}).

\section{Asymptotic analysis}\label{sec:asymptotics}

\begin{figure}[t!]
\includegraphics[width=6.0cm]{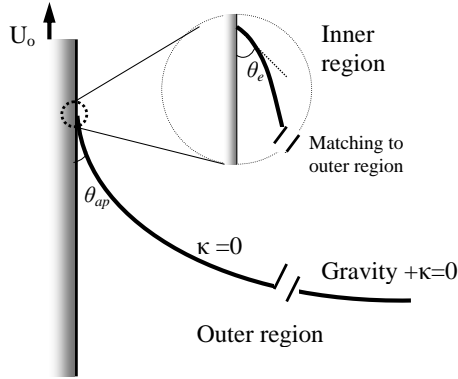}
\caption{Schematic diagram showing the different asymptotic regions for the case of a thin fiber. The inner region originates from a balance between viscosity and surface tension. It has a microscopic contact angle $\theta_e$. The outer region is a static meniscus joining a fiber with an apparent contact angle $\theta_{ap}$. When the fiber radius $r_0\ll \ell_c$, the outer profile is further separated into two regions~\cite{DFJ74}.}
\label{matching}
\end{figure}

We compute the shape of an axisymmetric meniscus on a fiber of radius $r_0$ using the method of matched asymptotic expansions. The interface is characterized by $h(x)$, as sketched in Fig.~\ref{fig.sketch}. The matching procedure is outlined schematically in Fig.~\ref{matching}. At small scales, the dominant balance is between viscosity $\eta$ and surface tension $\gamma$, and is characterized by the capillary number ${\rm Ca}$. Viscous effects can be neglected on large scales, for which the interface profile is that of a static meniscus. The problem is closed by matching the inner and outer solutions. The analysis provides the meniscus rise $\Delta$ as a function of ${\rm Ca}$ as well as the critical speed, both of which can be observed experimentally. We consider both large fiber radii ($r_0 \gg \ell_c$) and small fiber radii ($r_0 \ll \ell_c$). In all cases we take $r_0$ and $\ell_c$ to be macroscopic and much greater than the microscopic cutoff. Throughout the analysis, we scale all lengths by the capillary length, i.e. $\ell_c=1$.

\subsection{Inner solution: lubrication approximation}

To distinguish the solution $h$ in the inner region and the outer region, we denote $h_{in}(x)$ as the solution in inner region and $h_{out}(x)$ as the solution in outer region. The characteristic length scale for the inner solution comes from the cutoff of the viscous singularity, which here we take the slip length $\lambda$. Since typical interface curvatures turn out $\sim {\rm Ca}^{1/3}/\lambda$, as can be observed from the rescalings below, we can neglect the curvature contribution due to axisymmetry, which is of order $1/r_0$. Hence, for the inner solution we can follow the analysis by Eggers~\cite{E04b,E05}, which was originally derived for the flat plate. For completeness, we briefly summarize the analysis and the central results.

By restricting the analysis to small contact angles, $h_{in}'(0) = \theta_e \ll 1$, one can determine $h(x)$ from the lubrication approximation~\cite{ODB97}:

\begin{equation}
\label{lub}
h_{in}''' = \frac{3{\rm Ca}}{h_{in}^2 + 3\lambda h_{in}}.
\end{equation}
Since the slip length $\lambda$ is the only length scale, we rescale the solutions according to

\begin{equation}
\label{scal}
h_{in}(x) = 3\lambda H\left(\frac{x \theta_e }{3\lambda}\right),
\quad \xi = \frac{x \theta_e}{3\lambda}.
\end{equation}
Hence (\ref{scal}) reduces to

\begin{equation}
\label{simd}
H''' = \frac{\delta}{H^2 + H},
\end{equation}
where we introduced a reduced capillary number $\delta = 3{\rm Ca}/\theta_e^3$.
The boundary conditions are

\begin{equation}
H(\xi=0) = 0,
\end{equation}

\begin{equation}
H'(\xi=0) = 1
\end{equation}
and the asymptotic behavior that has to be matched to the outer solution.
Away from the contact line, where $H \gg 1$, (\ref{simd}) further reduces to

\begin{equation}
\label{sima}
y''' = \frac{1}{y^2},
\end{equation}
where we have put $H(\xi) = \delta^{1/3}y(\xi)$. This equation has an exact solution, whose properties have been summarized in \cite{DW97}. In parametric form, a solution with
$y(0)=0$ reads

\begin{eqnarray}
\label{par}
\left.\begin{array}{l}
\xi = \frac{2^{1/3}\pi Ai(s)}{\beta(\alpha Ai(s) + \beta Bi(s))} \\
y = \frac{1}{(\alpha Ai(s) + \beta Bi(s))^2}
                 \end{array}\right\}s\in [s_1 , \infty[,
\end{eqnarray}
where $Ai$ and $Bi$ are Airy functions~\cite{AS68}.
The limit $\xi\rightarrow 0$ corresponds to $s\rightarrow \infty$,
the opposite limit $\xi\rightarrow \infty$ to
$s\rightarrow s_1$, where $s_1$ is a root
of the denominator of (\ref{par}):
\begin{equation}
\label{s1}
\alpha Ai(s_1)+\beta Bi(s_1)=0.
\end{equation}
Since the solution extends to
$s=\infty$, $s_1$ has to be the largest root of (\ref{s1}).

The solution $y(\xi)$ is thus characterized by three parameters $\alpha$, $\beta$ and
$s_1$. Note that these are related according to (\ref{s1}), so that only two parameters are independent. The constant $\beta$ can be determined by
matching (\ref{par}), which is valid only for
$\xi \mbox{\ \raisebox{-.9ex}{$\stackrel{\textstyle >}{\sim}$}\ } 1$,
to a solution of (\ref{simd}), which includes the effect of the cutoff
and is thus valid down to the position $\xi=0$ of the contact line~\cite{E05}.
It was found that

\begin{equation}
\label{beta}
\beta^2=\pi\exp(-1/(3\delta))/2^{2/3} + O(\delta),
\end{equation}
which eliminates one of the two free parameters. The remaining parameter will be eliminated below by matching the large scale asymptotics of $y(\xi)$ the outer solution of the problem. For that, we only need the asymptotic behavior of $y(\xi)$ for large $\xi$, which reads:

\begin{equation}
\label{curv}
y(\xi) = \frac{1}{2} \kappa_y \xi^2 + b_y \xi + \mathcal{O}(1),
\end{equation}
where
\begin{equation}\label{eq:kb}
\kappa_y = \left(\frac{2^{1/6}\beta}{\pi Ai(s_1)}\right)^2, \quad
b_y = \frac{-2^{2/3}Ai'(s_1)}{Ai(s_1)}.
\end{equation}

\subsection{Outer solution: static meniscus}

At the scale of outer solution one can neglect viscous effects, and the profile is governed by surface tension and gravity. Thus equating the hydrostatic pressure and the capillary pressure gives

\begin{equation}\label{staticfiber}
\kappa=\Delta-x,
\end{equation}
where $\kappa$ is the curvature of the interface. Remind that we expressed all lengths in the capillary length $\ell_c=1$. The curvature can be expressed from the geometric relation

\begin{equation}\label{kafiber}
\kappa=\frac{h_{\rm out}^{''}}{(1+h_{\rm out}^{'2})^{3/2}}-\frac{1}{(r_0+h_{\rm out})(1+h_{\rm out}^{'2})^{1/2}}.
\end{equation}
The corresponding outer solution $h_{out}(x)$ is that of a meniscus of a liquid reservoir joining the fiber surface. The contact angle of the meniscus at the surface is denoted as the apparent contact angle, $\theta_{ap}$, since it refers to the apparent angle on the scale of the outer solution. The boundary conditions therefore are:

\begin{eqnarray}
h_{\rm out}(x=0)&=&0, \\
h^{'}_{\rm out}(x=0)&=&\theta_{ap}, \\
h^{'}_{\rm out}(x=\Delta)&=&\infty.
\end{eqnarray}
For the present analysis we require only the asymptotic behavior near the contact line, which is obtained by a Taylor expansion,

\begin{equation}\label{eq:out}
h_{\rm out} (x) = \theta_{ap} x + \frac{1}{2} \kappa_{ap} x^2 + \mathcal{O}\left(x^3 \right).
\end{equation}
Note that we consider small $\theta_{ap}$, since the inner solution is obtained in the lubrication limit.

In general, the governing equation (\ref{staticfiber}) cannot be solved analytically. In the following we will consider two extreme cases for which analytical solution can be obtained, namely the larger fiber radius case ($r_0\gg 1$) and the small fiber radius case ($r_0\ll 1$).

\subsubsection{Large fiber radius: $r_0 \gg 1$}

In the case where the fiber radius is much larger than the capillary length, the second term on the right hand side of (\ref{kafiber}) due to the curvature of the fiber can be neglected. Then (\ref{staticfiber}) can be written as

 \begin{equation}\label{outlar}
\frac{h_{\rm out}^{''}}{(1+h_{\rm out}^{'2})^{3/2}}=\Delta-x.
\end{equation}
Integrating (\ref{outlar}) once with respect to $x$, we obtain
\begin{equation}\label{outz2}
1-\frac{h_{\rm out}^{'}}{(1+h_{\rm out}^{'2})^{1/2}}=\frac{1}{2}(\Delta-x)^2,
\end{equation}
where the boundary condition $h_{\rm out}^{'}\rightarrow\infty$ at the position of the reservoir ($x = \Delta$) is used. Evaluating (\ref{outz2}) at the contact line position ($x = 0$) and using the geometrical connection $\sin\theta = h_{\rm out}^{'}/\sqrt{1+h_{\rm out}^{'2}}$, we end up with
\begin{eqnarray}
\Delta  &=&  \sqrt{2(1-\sin\theta_{ap})} \\
&\simeq&  \sqrt{2}(1-\theta_{ap}/2)
\end{eqnarray}
and (\ref{staticfiber}) immediately gives
\begin{equation}\label{eq:blabla}
\kappa_{ap} \simeq \sqrt{2}(1-\theta_{ap}/2).
\end{equation}

\subsubsection{Small fiber radius: $r_0 \ll 1$}

For thin fibers it has been shown that the outer region can be further divided into two subregions~\cite{DFJ74}, as has been sketched in Fig.~\ref{matching}. In the region far away from the fiber ($h\gg1$), the term due to the curvature of the fiber can be neglected. On the other hand, gravity can be neglected in the region close to the fiber ($h\ll1$), and the meniscus is determined by the balance between the two curvature terms in (\ref{kafiber}). The profile near the fiber is a classical zero curvature interface that can be expressed as

\begin{eqnarray}
h_{\rm out}(x)=\nonumber \\
r_0\left[\cosh\left(\frac{x}{r_0\cos\theta_{ap}}\right)+\sin\theta_{ap}\sinh\left(\frac{x}{r_0\cos\theta_{ap}}\right)-1\right]. \nonumber \\
\end{eqnarray}
In the following paragraphs we will match this small-scale part of the outer solution to the viscous inner solution. We therefore make a Taylor expansion, for small values of $\theta_{ap}$,

\begin{equation}\label{eq:fiberout}
h_{out}=\theta_{ap}x+\frac{1}{2r_0} x^2+\mathcal{O}(x^3).
\end{equation}

To express our results in terms of the meniscus rise $\Delta$, we quote the result obtained by James~\cite{DFJ74} in which the two subregions of the outer meniscus were matched:

\begin{equation}\label{del}
\Delta=r_0\left[\ln\left(\frac{4}{r_0(1+\sin\theta_{ap})}\right)-c\right],
\end{equation}
where c is Euler's constant (0.57721...).

\subsection{Matching}
We are now in a position to perform the matching between inner and outer solutions. First, we write the inner solution in terms of the original variables,

\begin{eqnarray}\label{eq:in}
h_{\rm in}(x) &=& \delta^{1/3} \left[ \frac{\kappa_y \theta_e^2 x^2}
{6\lambda}   + b_y \theta_e x +  \mathcal{O}\left(1 \right) \right].
\end{eqnarray}
Once more, we separately discuss the limits of large and small fiber radii.

\subsubsection{Large fiber radius: $r_0 \gg 1$}

Comparing the inner solution (\ref{eq:in}) to the outer solution (\ref{eq:out},\ref{eq:blabla}) one finds the matching conditions

\begin{eqnarray}\label{eq:match}
\theta_{ap}  &=& \delta^{1/3} b_y \theta_e, \\
 2- \theta_{ap} &=& \sqrt{2} \delta^{1/3} \frac{\kappa_y \theta_e^2 }
{3\lambda}.
\end{eqnarray}
Adding these two conditions leads to an equation for $s_1$ as a
function of $\delta$:

\begin{equation}
\label{eq:condition}
\frac{2/\theta_e}{\delta^{1/3}} + \frac{2^{2/3} Ai'(s_1)}{Ai(s_1)}  =
 \frac{2^{1/6}\exp[-1/(3\delta)] }{3\pi Ai^2(s_1) \lambda/\theta_e }.
\end{equation}
Once $s_1$ is known, one can compute the apparent contact angle

\begin{equation}\label{app}
\frac{\theta_{ap}}{\theta_e} = \frac{-2^{2/3}\delta^{1/3} Ai'(s_1)}{Ai(s_1)}.
\end{equation}

\begin{figure}[t!]
\includegraphics[width=7.0cm]{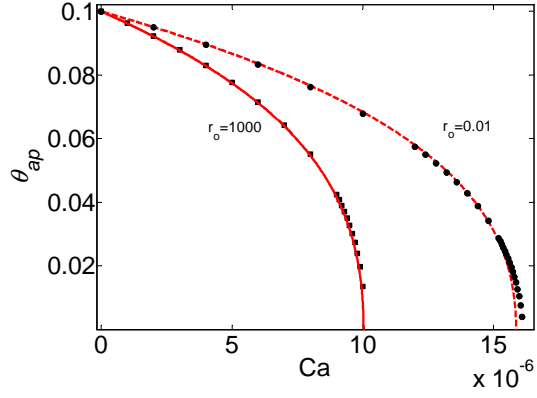}
\caption{(color online) Apparent contact angle $\theta_{ap}$ versus $\rm Ca$ ($\lambda = 10^{-8}$, $\theta_e = 0.1$ radian) for large radius ($r_0$ = 1000) and small radius ($r_0$ = 0.01). Curves: result from asymptotic matching, solid curve: $r_0 = 1000$, dashed curve: $r_0 = 0.01$; Symbols: numerical result, squares: $r_0 = 1000$, circles: $r_0 = 0.01$.}
\label{fig.theta}
\end{figure}

A typical result for $\theta_{ap}$ as a function of ${\rm Ca}$ is shown in Fig.~\ref{fig.theta} (solid curve: $r_0 = 1000$). At vanishing speed, one recovers the equilibrium contact angle ($\theta_e=0.1$ in this example). The apparent contact angle decreases for increasing speed, and tends to $\theta_{ap}=0$ at a critical value ${\rm Ca}_c$. The prediction from the matching compares very well to direct numerical solution of the problem, which will be discussed in the following section (solid squares). Of course, it is also possible to determine the critical speed directly from (\ref{eq:condition}), as shown in~\cite{E05}. The critical value $\delta_c$ is obtained when the Airy function takes its global maximum, $Ai'(s_1) = 0$,
corresponding to $s_{\rm max} = -1.088\cdots$. This gives a critical speed

\begin{equation}\label{delclarge}
\delta_c  = \frac{1}{3}\left[
\ln\left(
\frac{\delta_c^{1/3}\theta_e^2}{2^{5/6}3\pi (Ai(s_{\rm max}))^2 \lambda }
\right)
\right]^{-1}.
\end{equation}
Note that $\delta=3\rm Ca/ \theta_{e}^{3}$ and $Ai(s_{max})$=0.53566.... Physically, this corresponds to a vanishing apparent contact angle, as can be seen from (\ref{app}) since $Ai'(s_1)$=0. Indeed, this confirms the conjecture by Derjaguin and Levi~\cite{DL64} that the maximum speed is attained when $\theta_{ap}=0$.

\subsubsection{Small fiber radius: $r_0 \ll 1$}

We now perform a similar analysis for thin fibers by using the outer solution (\ref{eq:fiberout}), which was never worked out previously. Comparing this to the inner solution (\ref{eq:in}) one finds the matching conditions

\begin{equation}\label{sfcondition1}
\theta_{ap}=\delta^{1/3}b_y\theta_e,
\end{equation}

\begin{equation}\label{sfcondition2}
\frac{1}{r_0}=\delta^{1/3}\frac{\kappa_y\theta^2_e}{3\lambda}.
\end{equation}
The parameter $s_1$ can be solved as function of $\delta$ from (\ref{sfcondition2}). More explicitly, we can write (\ref{sfcondition2}) as

\begin{equation}\label{eq:condition2}
\frac{2/\theta_e}{\delta^{1/3}} = \frac{2^{2/3}r_0\exp[-1/(3\delta)]}{3\pi Ai^2(s_1) \lambda/\theta_e }.
\end{equation}
The apparent contact angle follows from (\ref{sfcondition1}). Since this condition is the same for both small fiber radius and large fiber radius, the explicit form of $\theta_{ap}$ is also given by eqn. (\ref{app}).

Once again, solutions of the matching conditions cease to exist at a critical speed, which occurs when the Airy function takes it global maximum, $Ai'(s_1) = 0$. In perfect analogy to the flat plate case, this corresponds to $\theta_{ap}=0$. The critical speed is given by

\begin{equation}\label{delc}
\delta_c  = \frac{1}{3}\left[
\ln\left(
\frac{r_0\delta_c^{1/3}\theta_e^2}{2^{1/3}3\pi (Ai(s_{\rm max}))^2 \lambda }
\right)
\right]^{-1}.
\end{equation}
This result has the same structure as (\ref{delclarge}), valid for $r_0\gg 1$. Apart from numerical coefficients, the main difference is that the fiber radius $r_0$ appears inside the logarithm as the relevant outer length scale; for the flat plate the outer scale is the capillary length.

This result is further illustrated in Fig.~\ref{fig.theta} showing $\theta_{ap}$ for a radius $r_0=10^{-2}$ (dashed curve). The curve is similar to that obtained for a plate of infinite radius, with a vanishing contact angle at the critical point. Note that this critical speed depends weakly (logarithmically) on the fiber radius, in agreement with prediction (\ref{delc}). In addition, there is a also a logarithmic dependence of $\delta_c$ on the equilibrium contact angle $\theta_e$. Let us emphasize that the validity of the asymptotic analysis requires $\lambda/\theta_e$ to be small. This means that, strictly speaking, we cannot deal with extremely small values of $\theta_e$.

It is instructive to compare our results with Voinov's formula \cite{V76}. The prediction by Voinov for $\delta_c$ has the same structure as ours, but the factor inside the logarithm is not precisely specified (a ratio between the macroscopic length scale and the microscopic scale). In fact the factor reflects the dependence on the specific geometry of the problem, that in our approach is determined by the matching of the inner region and the outer region. Naturally, the inner scale turns out to be the slip length, while the outer scale is the fiber radius or the capillary length. However, Voinov's formula misses details like the factors $\theta_e$, $\delta_c$ inside the logarithm. Also the resulting $\theta_{ap}$ vs $\rm Ca$ is a bit different from Voinov's formula, as was previously discussed in detail by Eggers \cite{E05}.

\section{Numerical solution}\label{sec:bifurcation}

We now perform a numerical analysis of the fiber withdrawal problem. This will confirm the validity of the asymptotics and extend the results to $r_0 \sim 1$. However, the main added value is that the numerical solution can determine the complete bifurcation diagrams of dewetting for arbitrary $r_0$. These contain steady state solutions above $\Delta_{\rm max}$ that serve as transients towards film deposition~\cite{SADF07,ZSE09}, and thus provide crucial additional information. Below we first develop a lubrication model that accounts for the axisymmetric nature of the flow. This quantitative correction with respect to the flat plate will turn out important for the bifurcation diagram. We then summarize the numerical results.

\subsection{Lubrication approximation on a fiber}

To formulate a hydrodynamic model for the axisymmetric meniscus on a fiber, we consider Stokes equations

\begin{eqnarray}
-\vec{\nabla} p+\eta \nabla^2 \vec{U}-\vec{\nabla} \Phi=0,
\end{eqnarray}
\begin{eqnarray}
\vec{\nabla} \cdot\vec{U}=0,
\end{eqnarray}
where $p$ is the pressure field in the liquid, $\eta$ is the viscosity of the liquid, $ \vec{U}$ is the velocity field in the frame comoving with the fiber, and $\Phi$ is the gravitational potential per unit volume in the liquid. Since the meniscus is axisymmetric, the velocity in azimuthal direction is zero. We consider small contact angle, $\theta_{e}\ll1$, thus the flow is mainly in the vertical $x$ direction, namely, the radial component of velocity is much smaller than the vertical component (i.e. $|U_r|\ll |U_x|$). The flow is solved with a no-stress condition at the interface, is located at $r=r_0+h$, and reads (in the frame of the fiber)

\begin{eqnarray}
\eta\left( \frac{\partial U_r}{\partial x}+\frac{\partial U_x}{\partial r}\right)_{r=r_0+h}\approx\eta\frac{\partial U_x}{\partial r}|_{r=r_0+h}=0.
\end{eqnarray}
At the fiber surface, $r=r_0$, we apply a Navier slip boundary condition

\begin{eqnarray}
U_x|_{r=r_0}=\lambda\frac{\partial U_x}{\partial r}|_{r=r_0}.
\end{eqnarray}
The axial (vertical) component of the velocity field then becomes

\begin{eqnarray}\label{uz}
U_x=\frac{1}{2\eta}\frac{\partial (p+\Phi)}{\partial x} \nonumber \\
\times\left[\frac{r^2-r_0^2}{2}-(r_0+h)^2\ln \left(\frac{r}{r_0}\right)-\lambda \left(2h+\frac{h^2}{r_0}\right)\right]
\end{eqnarray}
For thin films $h/r_0 \ll 1$ this reduces to the usual parabolic Poiseuille profile, but quantitative corrections appear when $h/r_0 \sim 1$.

The lubrication equation is obtained by imposing a zero flux condition in the frame of the reservoir

\begin{equation}
\int_{r_0}^{r_0+h} (U_x+U_0) rdr=0.
\end{equation}
With this, (\ref{uz}) can then be simplified as

\begin{equation}
\frac{\partial (p+\Phi)}{\partial x}=\frac{3\eta U_0 f(d)}{h[h+3\lambda(1+d/2)f(d)]},
\end{equation}
where we introduced $d$=$h/r_0$ and

\begin{equation}
f(d)=\frac{8d^3(2+d)}{3[4(1+d)^4\ln(1+d)-d(2+d)(2+6d+3d^2)]}.
\end{equation}
This function is a correction factor with respect to the flat plate ($d=0$), and has the property $f(0)=1$. Finally, we replace the pressure by the Young-Laplace equation,

\begin{equation}
p-p_0=-\gamma\kappa,
\end{equation}
where $\kappa$ is the curvature of the interface given by (\ref{kafiber}). This gives the lubrication equation on a fiber

\begin{equation}\label{lubeq}
\frac{\partial \kappa}{\partial x}=\frac{3{\rm Ca}f(d) }{h[h+3\lambda(1+d/2)f(d)]}-1.
\end{equation}
Note that once again all lengths are scaled by the capillary length. For $d=h/r_0 \ll 1$, we recover the usual lubrication equation since $f(0)=1$.

\subsection{Results}

\subsubsection{Critical speed}

The above lubrication equation (\ref{lubeq}) is solved numerically with boundary conditions

\begin{eqnarray}
h(0) &=&0 ,\\
h'(0) &=& \theta_e,
 \end{eqnarray}
imposed at the contact line and

\begin{eqnarray}
h'(\Delta) &=& \infty,\\
\kappa(\Delta) &=& 0,
\end{eqnarray}
at the reservoir. We varied $r_0$ and $\lambda$ and determined the meniscus as a function of ${\rm Ca}$.

\begin{figure}[t!]
\includegraphics[width=8.0cm]{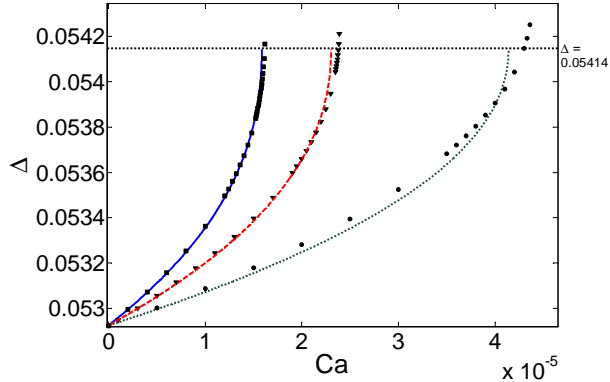}
\caption{(color online) Height of meniscus $\Delta$ versus speed for different slip lengths ($r_0=10^{-2}$, $\theta_e$=0.1 radian). Curves are results from asymptotic matching, solid curve (the most left curve): $\lambda=10^{-8}$, dashed curve: $\lambda=10^{-7}$, dotted curve (the most right curve): $\lambda=10^{-6}$. Symbols are the corresponding numerical results. As the slip length gets smaller, the agreement between numerics and asymptotic matching becomes better. The horizontal dotted line indicates the maximum height of meniscus $\Delta=0.05414$ calculated by (\ref{del}) with $\theta_{ap}=0$.}
\label{delta_ca_slipl}
\end{figure}

Figure~\ref{delta_ca_slipl} shows the meniscus rise $\Delta$ as a function of ${\rm Ca}$ on a fiber of radius $r_0=10^{-2}$. Different symbols correspond to different values of the slip length. In all cases we find a critical $\rm Ca_c$ above which solutions cease to exist. This indeed occurs close to $\Delta_{\rm max}$ corresponding to a vanishing $\theta_{ap}$, which is indicated by the horizontal dotted line. The curves are the predictions from the matched asymptotics, showing a good agreement with the numerical solutions. In particular, one observes convergence as the slip length is reduced from  $\lambda$ = $10^{-6}$, $10^{-7}$ to $10^{-8}$. This is because the separation of the two length scales $\lambda$ and $r_0$ is enhanced, which improves the validity of the matching asymptotic expansion. The same results were previously reported in Fig.~\ref{fig.theta}, expressed in terms of $\theta_{ap}$ rather than $\Delta$.

\begin{figure}[t!]
\includegraphics[width=7.0cm]{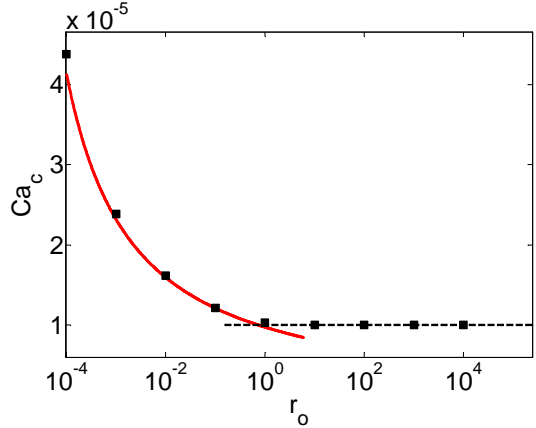}
\caption{(color online) Critical capillary number $\rm Ca_c$ versus fiber radius $r_0$ ($\theta_e$=0.1 radian, $\lambda$=10$^{-8}$).  Squares: numerical results; Curve: Result from asymptotic matching for small fiber radius [Eq. (\ref{delc})]; Dotted line: Result from asymptotic matching for large fiber radius [Eq. (\ref{delclarge})]. }
\label{fig.radius}
\end{figure}

It is interesting to show how the critical speed ${\rm Ca}_c$ depends on the fiber radius $r_0$. The numerical results are plotted as squares in Fig.~\ref{fig.radius}. In agreement with the asymptotic analysis one observes two regimes. At small radii, $r_0\ll 1$, the critical speed depends logarithmically on the radius. The solid red line is the asymptotic result (\ref{delc}). For large radii the speed approaches the value of the flat plate (\ref{delclarge}), indicated as dashed black line. Indeed, the cross-over occurs for fibers with a radius that is comparable to the capillary length $r_0\sim 1$.

\begin{figure*}[t!]
\includegraphics[width=18.5cm]{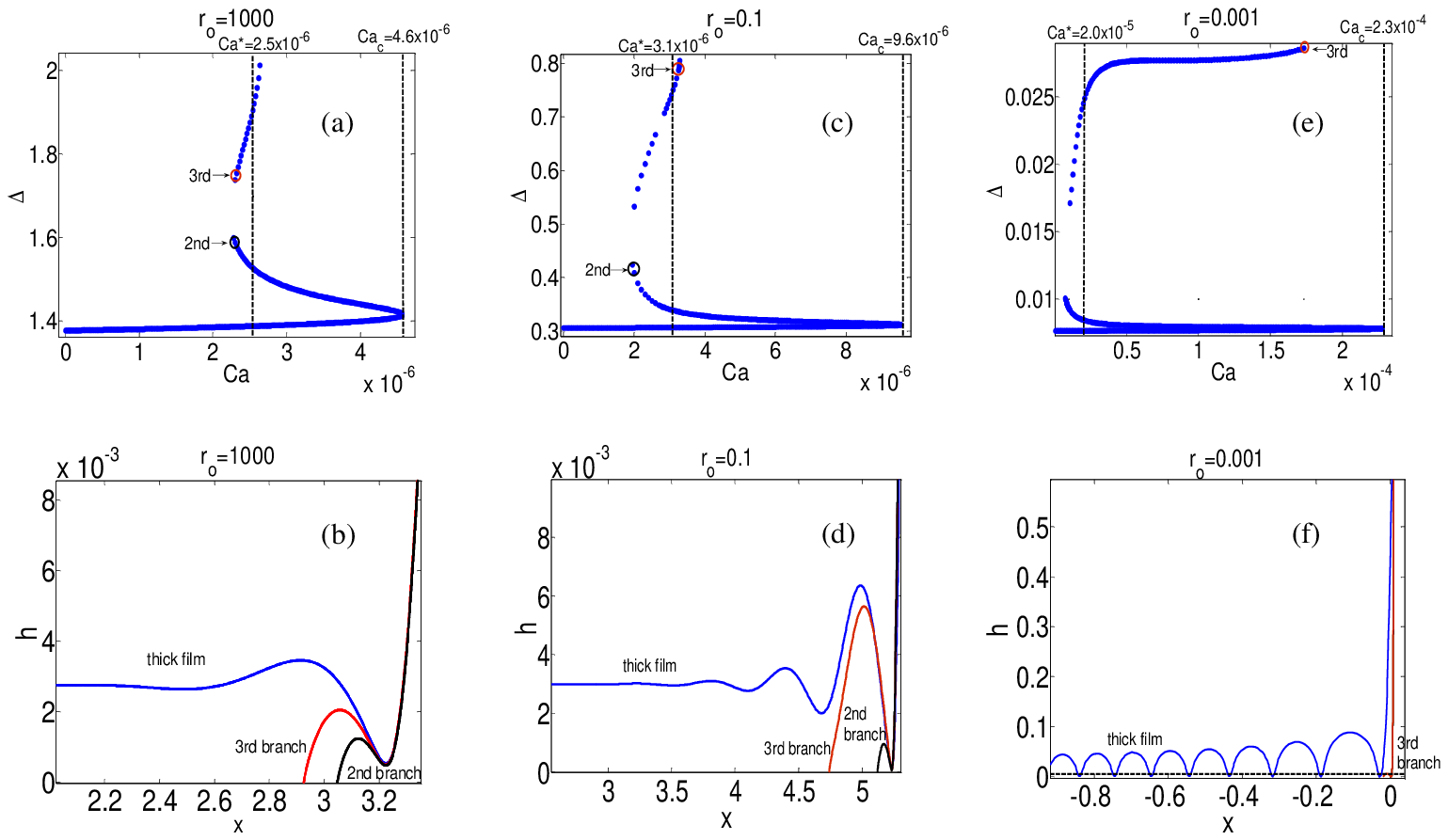}
\caption{(color online) Bifurcation diagrams of steady solutions. Panels (a), (c), (e): bifurcation diagrams for $r_0$=1000, 0.1 and 0.001 respectively. All curves correspond to $\theta_e=0.05$ and  $\lambda$=10$^{-5}$. Panels (b), (d), (f): interface profiles for $r_0$=1000, 0.1 and 0.001 respectively. We report profiles from the 2nd branch and 3rd branch, corresponding to solutions indicated in the bifurcation diagram by large circles. The ``thick film'' solutions correspond to profiles without contact line, or $\Delta \rightarrow \infty$, for which we define $\rm Ca=\rm Ca^*$. The thickness of the film for $r_0=0.001$ is shown by the horizontal line just above $x$ axis in (f).}
\label{all2}
\end{figure*}

\subsubsection{Meniscus rise: bifurcation diagram}

The results shown in Fig.~\ref{delta_ca_slipl} represent only the lowest branch of solutions of a more complete bifurcation diagram. Indeed, one can identify solutions with $\Delta$ extending to arbitrary height above the meniscus, which are all characterized by ${\rm Ca} < {\rm Ca_c}$. These are summarized in Figs.~\ref{all2}ace for different fiber radii (all curves correspond to $\theta_e$=0.05 and $\lambda$=$10^{-5}$). For $r_0=1000$ (Fig. \ref{all2}a), we see after reaching $\rm Ca_c$, the curve turns back to $\rm Ca < \rm Ca_c$ but with solutions of increasing $\Delta$. We refer to these solutions as the second branch, which is known to be unstable~\cite{SADF07}. Further upwards we observe a series of bifurcations to higher branches, oscillating around a characteristic value $\rm Ca^*$. Typical meniscus profiles are shown in Fig.~\ref{all2}b -- in order to compare the profiles we have shifted the positions of the contact line such that the baths collapse. Following the bifurcation diagram, the profiles evolve to a film solution for which the contact line has moved to arbitrary height above the meniscus, i.e. $\Delta \rightarrow \infty$. This film solution (shown in Fig.~\ref{all2}b) is not the Landau-Levich-Derjaguin film, but corresponds to the new class of ``thick film'' solutions identified in~\cite{SZAFE08}.

We then decrease the fiber radius to $r_0=0.1$,  as shown in Figs.~\ref{all2}cd. We find that $\rm Ca_c$ increases almost by a factor 2 with respect to the large radius. By contrast, $\rm Ca^*$ corresponding to the thick film increases only by a small amount. As a result the values of $\rm Ca_c$ and $\rm Ca^*$ have become more separated. Also, the corresponding meniscus profiles display more structure. The thick film exhibits much stronger oscillations before joining the reservoir. These trends becomes more dramatic up further decreasing the radius $r_0=0.001$ (Figs.~\ref{all2}ed). The difference between $\rm Ca_c$ and $\rm Ca^*$ is very pronounced, and $\Delta$ changes much more dramatically for the 3rd branch solutions. In this sense, the bifurcation diagram has a very different structure from those of large fiber radius. Interestingly, there still exists a thick film solution matching to the bath, but the profile displays many oscillations (Fig.~\ref{all2}f). These oscillations decay only very slowly when moving further away from the bath -- the asymptotic thickness of the film is indicated by the horizontal line just above $x$ axis. While for the flat plate the thick film solutions have been observed experimentally~\cite{SZAFE08}, we expect the oscillatory solution obtained for small radii to be unstable and of no physical relevance.

\begin{figure}[t!]
\includegraphics[width=8.5cm]{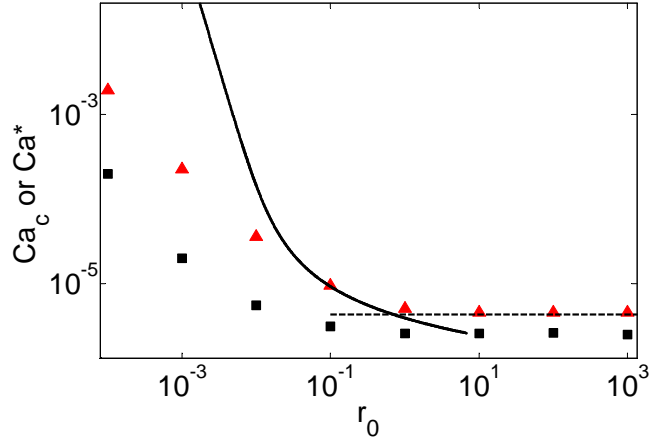}
\caption{$\rm Ca_c$ and $\rm Ca^*$ as function of $r_0$ for the same parameters used in Fig.~\ref{all2}. Triangles: numerical results for $\rm Ca_c$; Squares: numerical results for $\rm Ca^*$; Curve: Result for $\rm Ca_c$ from asymptotic matching for small fiber radius [Eq. (\ref{delc})]; Dotted line: Result for $\rm Ca_c$ from asymptotic matching for large fiber radius [Eq. (\ref{delclarge})].}
\label{ca_star}
\end{figure}

For completeness, we report the values of ${\rm Ca}_c$ and ${\rm Ca}^*$ for different radii in a separate graph (Fig.~\ref{ca_star}). Note that the theoretical curve for $\rm Ca_c$ deviates from the numerical results as early as $r_0 \lesssim 10^{-2}$. The reason is that here we use a realistic value for the slip length $\lambda = 10^{-5}$ (corresponding to $\sim 10$nm), instead of $\lambda = 10^{-8}$ used in Fig. \ref{fig.radius}. Clearly, the scale separation required for the asymptotic analysis starts to break down when the ratio $\lambda/r_0$ is no longer very small.

\section{Discussion}\label{sec:discussion}

We investigated the steady-state profiles of axisymmetric menisci on a fiber that is withdrawn from a viscous liquid. The main motivation for this work was the mixed experimental observations on the transition to film deposition obtained for fibers, large cylinders and plates. Sedev \& Petrov~\cite{SP91} found that the maximum steady profile has a meniscus rise identical to a perfectly wetting liquid at equilibrium, suggesting a vanishing apparent contact angle $\theta_{ap}$. Other experiments found that steady-state solutions disappeared at a nonzero $\theta_{ap}$~\cite{MRQG07,SDAF06}, although the critical point could be accessed during transients~\cite{DFSA07}. Our present calculations show that steady solutions always cease to exist at $\theta_{ap}=0$, independent of the fiber radius. In addition, stability arguments put forward in \cite{GR03,E05} suggest that all solutions of the lowest branch are perfectly stable up to the maximum speed, consistent with a saddle-node bifurcation~\cite{SADF07}. In that sense, our results do not provide an explanation why experimentally it is practically impossible to achieve steady menisci closer to $\theta_{ap}=0$. The main effect that was not taken into account in our calculations is contact angle hysteresis due to heterogeneity of the substrate~\cite{JG84}, which was previously suggested to affects the details of the transition~\cite{GR01}. It has remained a challenge, however, to incorporate this into a full hydrodynamic description of moving contact lines. Essentially one has to modify the boundary condition by imposing a time-dependent microscopic contact angle at the moving contact line.

The bifurcation diagrams calculated in the second part of the paper, however, do provide a new experimental perspective on the dynamics of film deposition. As shown in~\cite{DFSA07}, such bifurcation diagrams may be probed experimentally as transient states during entrainment. Namely, for ${\rm Ca}>{\rm Ca}_c$ the evolution of the meniscus exactly follows the bifurcation diagram when plotting $\Delta$ versus the {\em relative} contact line velocity with respect to the solid. For very large fiber radius the profiles with large capillary rise are smoothly connected to the bath by a film that only displays a small ``dimple'' close to the bath. These dimple solutions have indeed been observed experimentally when plates are withdrawn with speeds above the critical speed. By contrast, for small fiber radii these solutions exhibit very strong oscillations (Fig.~\ref{all2}) and we expect these solutions to be very unstable. In that case another dynamical mode must appear in order to deposit a liquid film -- for example, one could think of the classical dewetting rim at the contact line connected to a Landau-Levich film~\cite{RBR91}. A further investigation of these transients above the critical speed, in particular for different radii, should give a more complete picture of the forced wetting transition.

\begin{acknowledgments}
We are grateful to Jens Eggers and Bruno Andreotti for discussions and support. TSC acknowledges financial support by the FP7 Marie Curie Initial Training Network ``Surface Physics for Advanced Manufacturing'' project ITN 215723.
\end{acknowledgments}

\bibliography{all_ref}

\end{document}